\documentstyle[12pt,epsfig,palatino]{article}

\newcommand{\mysection}{\setcounter{equation}{0}\section}

\renewcommand{\thefootnote}{\arabic{footnote}}
\newcommand{\ct}{\mbox{$\cos^2\theta^*$}}
\newcommand{\cf}{\mbox{$\cos^4\theta^*$}}
\newcommand{\co}{\mbox{$\cos\theta^*$}}
\newcommand{\cth}{\mbox{$\cos^3\theta^*$}}
\begin{document}
\baselineskip 22pt
\vspace*{-1in}
\renewcommand{\thefootnote}{\fnsymbol{footnote}}
\begin{flushright}
SINP/TNP/05-16\\
{\tt hep-ph/0507250}
\end{flushright}
\vskip 65pt
\begin{center}
{\Large \bf Angular distribution of Drell-Yan process at hadron
colliders to NLO-QCD in models of TeV scale gravity}\\
\vspace{8mm}
{\bf
Prakash Mathews$^a$
\footnote{prakash.mathews@saha.ac.in}, 
V. Ravindran$^b$
\footnote{ravindra@mri.ernet.in}
}\\
\end{center}
\vspace{10pt}
\begin{flushleft}
{\it
a) 
Saha Institute of Nuclear Physics, 1/AF Bidhan Nagar,
Kolkata 700 064, India.\\

b) Harish-Chandra Research Institute,
 Chhatnag Road, Jhunsi, Allahabad, India.\\

}
\end{flushleft}

\vspace{10pt}
\begin{center}
{\bf ABSTRACT}
\end{center}
\vskip12pt
In TeV scale gravity models, for dilepton production at hadron colliders, 
we present the NLO-QCD corrections for the double differential cross 
section in the invariant mass and scattering angle.  For both ADD and 
RS models, the quantitative impact of QCD corrections for extra dimension 
searches at LHC and Tevatron are investigated.  We present the K-factors 
for both ADD and RS models at LHC and Tevatron.  Inclusion of QCD 
corrections to NLO stabilises the cross section with respect to scale 
variations.  

\vfill
\clearpage

\setcounter{page}{1}
\pagestyle{plain}

\mysection{Introduction}

Extra dimension scenarios are now essential part of the studies of 
physics beyond the Standard Model (SM).  They provide an alternate 
view of the hierarchy between the electroweak and the Planck 
scale.  Some of these extra dimension models invoke the brane world 
scenarios to hide the extra spacial dimensions from current observation.  
Two such models that are phenomenologically widely studied are the 
Arkani-Hamed, Dimopoulos and Dvali (ADD) \cite{add} and the 
Randall-Sundrum (RS) \cite{rs} models.

In the ADD case the compactified extra dimensions could be large and the 
large volume of the compactified extra spacial dimension would account for 
the dilution of gravity in 4-dimensions and hence the hierarchy.  In this 
model, new physics can appear at a mass scale of the order of a TeV.  A 
viable mechanism to hide the extra spacial dimension, is to introduce a 
3-brane with negligible tension and localise the SM particles on it.  Only 
the graviton is allowed to propagate the full $4+d$ dimensional space 
time.  As a consequence of these assumptions, it follows from Gauss Law 
that the effective Planck scale $M_P$ in 4-dimension is related to the 
$4+d$ dimensional fundamental scale $M_S$ through the volume of the 
compactified extra dimensions \cite{add}.  The extra dimensions 
are compactified on a torus of common circumference $R$.  The number of 
extra spacial dimension possible is $d>2$ from current experimental limits 
on deviation from inverse square law \cite{expt}.  The space time is 
factorisable and the 4-dimensional spectrum consists of the SM confined to 
4-dimensions and a tower of Kaluza-Klien (KK) modes of the graviton propagating 
the full $4+d$ dimensional space time. 

The interaction of the KK modes $h_{\mu\nu}^{(\vec n)}$ with the SM fields 
localised on the 3-brane is given by
\begin{eqnarray}
{\cal L}_{int} \sim - \frac{1}{M_P} \sum_{\vec n=0}^\infty T^{\mu\nu} (x) 
                      h_{\mu\nu}^{(\vec n)} (x) ~,
\end{eqnarray}
where $T^{\mu\nu}$ is the energy-momentum tensor of the SM fields on the 
3-brane.  The zero mode corresponds to the usual 4-dimensional massless 
graviton.  The KK modes are all $M_P$ suppressed but the high multiplicity 
could lead to observable effects at present and future colliders.  
The Feynman rules are given in \cite{grw,hlz}.  
 
In the RS model there is only one extra spacial dimension and the extra 
dimension is compactified to a circle of circumference $2 L$ and further 
orbifolded by identifying points related by $y \to -y$.  Two branes are 
placed at orbifold fixed points, $y=0$ with positive tension called the 
Planck brane and a second brane at $y=L$ with negative tension called the 
TeV brane.  For a special choice of parameters, it turns out that the 
5-dimensional Einstein equations have a warped solution for $0<y<L$ with 
metric $g_{\mu\nu} (x^\rho,y)=\exp(-2 k y) ~\eta_{\mu\nu}$, $g_{\mu y}=0$ 
and $g_{y y}=1$.  This space is not factorisable and has a constant negative 
curvature--- $AdS_5$ space-time.  $k$ is the curvature of the $AdS_5$ 
space-time and $\eta_{\mu\nu}$ is the usual 4-dimensional flat Minkowski 
metric.  In this model the mass scales vary with $y$ according to the 
exponential warp factor.  If gravity originates on the brane at $y=0$, 
TeV scales can be generated on the brane at $y=L$ for $k L \sim 10$.
The apparent hierarchy is generated by the exponential warp 
factor and no additional large hierarchies appear.  The size of the 
extra dimension is of the order of $M_P^{-1}$.  Further it has been 
showed that \cite{gw} the value of $k L$ can be stabilised without 
fine tuning by minimising the potential for the modulus field 
which describes the relative motion of the two branes.  In the RS 
model graviton and the modulus field can propagate the full
5-dimensional space time while the SM is confined to the TeV brane.
The 4-dimensional spectrum contains the KK modes, the zero mode is
$M_P$ suppressed while the excited modes are massive and are only TeV 
suppressed.  The mass gap of the KK modes is determined by the difference 
of the successive zeros of the Bessel function $J_1 (x)$ and the scale
$m_0=k ~e^{-\pi k L}$.  
As in the ADD case the phenomenology of the RS model concerns
the effect of massive KK modes of the graviton, though the spectrum
of the KK mode is quite different.  

In the RS model the massive KK modes $h_{\mu\nu}^{(n)} (x)$ interacts
with the SM fields 
\begin{eqnarray}
{\cal L}_{int} \sim - \frac{1}{M_P} T^{\mu\nu} (x) h_{\mu\nu}^{(0)} (x) 
                 - \frac{e^{\pi k L}}{M_P} \sum_{n=1}^\infty T^{\mu\nu} (x)
                   h_{\mu\nu}^{(n)} (x) ~,
\end{eqnarray}
where $T^{\mu\nu}$ is the energy-momentum tensor of the SM fields on the 
3-brane at $y=L$.  The masses of $h_{\mu\nu}^{(n)} (x)$ are given by $M_n=x_n 
~k ~e^{-\pi k L}$, where $x_n$ are the zeros of the Bessel function $J_1 
(x)$.  In the RS model there are two parameters which are $c_0=k/M_P$, the 
effective coupling 
\footnote{An alternate definition is ${\bar c}_0=k/{\overline M}_P$, where 
${\overline M}_P=M_P/\sqrt{8 \pi}$, hence ${\bar c}_0 ~= ~c_0 ~\sqrt{8 \pi}$.}
and $M_1$ the mass of the first KK mode.  Expect for 
an overall warp factor the Feynman rule of RS is the same as those of the 
ADD model.  

Next to leading order (NLO) QCD corrections have been recently calculated 
in the ADD case for $e^+ e^- \to $ hadrons \cite{us} and various distributions 
of invariant lepton pair production at both LHC and Tevatron \cite{us1}.  
This was 
further extended to the RS case \cite{us2}.  In this paper, for the 
ADD and RS models, we consider the un-integrated distribution with
respect to $\cos \theta^*$ to NLO in QCD, where $\theta^*$ is the 
scattering angle of the lepton with an initial hadron in the {\it c.o.m} 
frame of the 
lepton pair.  This is particularly important in the dilepton production 
case to achieve maximum sensitivity to the model parameters, as $\cos 
\theta^*$ integrated cross section is independent of the interference 
between SM and gravity \cite{us1}.  To leading order (LO), this double 
differential $d \sigma/d Q/d\cos \theta^*$ was analysed in \cite{cl}.  
At hadron colliders the NLO-QCD corrections are important especially 
in models of extra dimension as gluon-gluon subprocess contributes at 
the same LO as quark-antiquark subprocess.  D\O ~Collaboration recently 
reported searches for large extra dimensions in the dimuon channel for 
the double differential cross section \cite{d0_add1}, this updates the 
earlier Run-I results \cite{d0_add}.  The first direct search of the 
RS KK modes using the dileptons have been reported by D\O ~Collaboration
\cite{d0_rs}.

Rest of the paper is organised as follows:
In section 2 we evaluate the NLO coefficient functions to the subprocess 
that contribute to the double differential cross section $d \sigma /dQ 
/d \cos \theta^*$.  Finally in section 3, we discuss the impact of the 
NLO results.

\mysection{\boldmath Drell-Yan $\cos \theta^* $ distribution}

We consider the Drell-Yan process and study the double differential
cross section with respect to the invariant mass of the final lepton
pair and $\cos \theta^*$ the cosine of the angle between the final 
state lepton momenta and the initial state hadron in the $c.o.m$ frame 
of the lepton pair \footnote{An alternate definition of the angle has 
been considered in \cite{kod} to study the lepton helicity distribution 
in polarised Drell-Yan process.}.  The relevant kinematical formulation 
is detailed in \cite{us1}.  In the QCD improved parton model, the hadronic 
cross section 
can be expressed in terms of partonic cross sections convoluted with 
appropriate parton distribution functions. The coefficient functions 
to NLO in QCD are evaluated for both ADD and RS models.  The difference
between the two models depend on the spectrum of the KK modes and hence 
summation of the KK modes that contribute to the dilepton production leads
to different results \cite{us1,us2}.  

The hadronic part involves the computation of various processes
that contribute to $Q$ or $X_F$ or rapidity distributions that 
are presented in the reference \cite{us1}.  The angular distributions 
which are "odd" in $\cos \theta^*$ come mainly from the interferences
terms.  The non-vanishing odd contribution in the standard model sector 
comes from the interference of photon mediated processes 
with $Z$-boson mediated processes.  We also find that non-vanishing 
odd contributions come from the interference of standard model diagrams
with the graviton exchange diagrams.  These inference diagrams are absent 
in the computation of $Q,X_F$ and rapidity distributions where only even 
functions of  $\cos \theta^*$ contribute.  We have regularised all the 
divergences using dimensional regularisation.  The mass singularities are 
removed by the mass factorisation, for details refer to \cite{us1}.

We first present the angular distribution which is "even" in $\co$.
\begin{eqnarray}
\label{eq2.18}
2 S{d \sigma_{e}^{P_1P_2} \over dQ^2 d\cos \theta^*}\!\!\!&=&
\sum_q{\cal F}_{SM,q} \int_0^1~ {d x_1}~ \int_0^1 
~{dx_2}~ \int_0^1~ dz~ \delta(\tau-z x_1 x_2)
\nonumber\\[2ex]&&
\times \Bigg[ H_{q \bar q}(x_1,x_2,\mu_F^2) \Big(
\Delta_{q \bar q}^{(0),\gamma/Z}(z,Q^2,\mu_F^2)
 +a_s \Delta_{q \bar q}^{(1),\gamma/Z}(z,Q^2,\mu_F^2)\Big)
\nonumber\\[2ex] &&
+H_{q g}(x_1,x_2,\mu_F^2)
 a_s \Delta_{q g}^{(1),\gamma/Z}(z,\mu_F^2) 
\nonumber\\[2ex] &&
+H_{g q}(x_1,x_2,\mu_F^2)
 a_s \Delta_{g q}^{(1),\gamma/Z}(z,\mu_F^2) \Bigg]
\nonumber\\[2ex]&&
+\sum_q{\cal F}_{SMGR,q} \int_0^1~ {d x_1}~ \int_0^1
~{dx_2}~ \int_0^1~ dz~ \delta(\tau-z x_1 x_2)
\nonumber\\[2ex]&&
\times \Bigg[ H_{q \bar q}(x_1,x_2,\mu_F^2) \Big(
\Delta_{q \bar q}^{(0),G\gamma/Z}(z,Q^2,\mu_F^2)
 +a_s \Delta_{q \bar q}^{(1),G\gamma/Z}(z,Q^2,\mu_F^2)\Big)
\nonumber\\[2ex] &&
+H_{q g}(x_1,x_2,\mu_F^2)
 a_s \Delta_{q g}^{(1),G\gamma/Z}(z,\mu_F^2)
\nonumber\\[2ex]&&
+H_{g q}(x_1,x_2,\mu_F^2)
 a_s \Delta_{g q}^{(1),G\gamma/Z}(z,\mu_F^2) \Bigg]
\nonumber\\[2ex]
&&+\sum_q{\cal F}_{GR} \int_0^1~ {d x_1 }~ \int_0^1 
~{dx_2}~ \int_0^1~ dz~ \delta(\tau-z x_1 x_2)
\nonumber\\[2ex]&&
\times \Bigg[ H_{q \bar q}(x_1,x_2,\mu_F^2) \Big(
\Delta_{q \bar q}^{(0),G}(z,Q^2,\mu_F^2)
 +a_s \Delta_{q \bar q}^{(1),G}(z,Q^2,\mu_F^2)\Big)
\nonumber\\[2ex] &&
+ H_{q g}(x_1,x_2,\mu_F^2)
 a_s \Delta_{q g}^{(1),G}(z,Q^2,\mu_F^2) 
\nonumber\\[2ex] &&
+H_{g q}(x_1,x_2,\mu_F^2)
 a_s \Delta_{g q}^{(1),G}(z,Q^2,\mu_F^2) 
\nonumber\\[2ex]&&
+ H_{g g}(x_1,x_2,\mu_F^2) \Big(
\Delta_{g g}^{(0),G}(z,Q^2,\mu_F^2)
 +a_s \Delta_{g g}^{(1),G}(z,Q^2,\mu_F^2)\Big) \Bigg]\,,
\nonumber\\[2ex]
\end{eqnarray}
where $H_{a b}(x_1,x_2,\mu_F^2)$ are the renormalised partonic distributions
and $\Delta_{a b}(z,Q^2,\mu_F^2)$ are the coefficient function corresponding
to various subprocess cross section to NLO in QCD.
The factors ${\cal F}_{SM,q},{\cal F}_{GR}$ correspond to pure SM and gravity 
(GR) part respectively and are given in \cite{us1}, the factor that corresponds
to the interference of SM and gravity is 
\begin{eqnarray}
{\cal F}_{SMGR,q}&=&{\alpha \kappa^2 Q^2 \over 4 \pi }|{\cal D}(Q^2)|
\Bigg[{Q^2 (Q^2-M_Z^2) \over
\left((Q^2-M_Z^2)^2+M_Z^2 \Gamma_Z^2\right)c_w^2 s_w^2}
g_q^A g_e^A \Bigg] \, ,
\end{eqnarray}
where $\alpha$ is the fine structure constant, $\kappa=\sqrt{16 \pi}/M_P$.
The summation of the KK modes leads to ${\cal D}(Q^2)$ for ADD and RS case 
has been given in \cite{us1,us2}.

The leading order results read
\begin{eqnarray}
\Delta^{(0),\gamma/Z}_{q\bar q}&=&
{2 \pi \over N} \delta(1-z)\Bigg[{3 \over 8} \left(1 + \ct\right)
\Bigg]\, ,
\nonumber \\[2ex]
\Delta^{(0),G\gamma/Z}_{q \bar q}&=& { \pi \over 8 N} \delta(1-z)
\Bigg[-1+3 \ct\Bigg] \,,
\nonumber\\[2ex]
\Delta^{(0),G}_{q \bar q}&=&{\pi \over 8 N} \delta(1-z)
\Bigg[{5\over 8} (1 -3 \ct +4 \cf)\Bigg]\, ,
\nonumber \\[2ex]
\Delta^{(0),G}_{gg}&=&{\pi \over 2 (N^2-1)}  \delta(1-z)
\Bigg[{5 \over 8} (1-\cf)\Bigg]\, ,
\label{eq2.23}
\end{eqnarray}
and the next to leading order results read
\begin{eqnarray}
\Delta^{(1)\gamma/Z}_{q \bar q}&=&
\left({2 \pi \over N}\right) 4~ C_F \Bigg\{
\Bigg[ \Big(-4+2 \zeta(2)\Big)\delta(1-z)+
2 {1 \over (1-z)_+} \ln{\left(Q^2 \over \mu_F^2\right)}
\nonumber \\[2ex] &&
+4 \left({\ln(1-z) \over 1-z}\right)_+
+{3 \over 2} \delta(1-z)
\ln\left({Q^2 \over \mu_F^2}\right)
-(1+z) \ln\left({Q^2 (1-z)^2 \over \mu_F^2 z}\right)
\nonumber \\[2ex]&&
-2 {\ln(z) \over 1-z} \Bigg] \Bigg({3 \over 8} (1+\ct)\Bigg)
+
\Big[1-z\Big] {3 \over 8} \Bigg(1-3 \ct \Bigg) \Bigg\} \, ,
\nonumber \\[2ex]
\Delta^{(1)\gamma/Z}_{q (\bar q)g}&=&
\left({2 \pi \over N}\right) T_F\Bigg\{
\Bigg[ -4 z \log(z)+ 2 (1-2 z+2 z^2) \ln\left({Q^2 (1-z)^2 \over 
\mu_F^2 z}\right) -7 z^2 \Bigg]
\nonumber \\[2ex]
&&\times \Bigg({3 \over 8} (1+\ct)\Bigg)
+\Bigg[ {15 \over 8} +{3 \over 4} z +3 z \log(z) \Bigg]
\nonumber \\[2ex]
&&+\Bigg[-{33 \over 8} +{27 \over 4} z -3 z \log(z) \Bigg]
\ct \Bigg\} \, ,
\nonumber \\[2ex]
\Delta^{(1)\gamma/Z}_{g q (\bar q)}&=&
\left({2 \pi \over N}\right) T_F \Bigg\{
\Bigg[ -4 z \log(z)+ 2 (1-2 z+2 z^2) \ln\left({Q^2 (1-z)^2 \over \mu_F^2 z}\right)
\Bigg]
\nonumber \\[2ex]
&&\times \Bigg({3 \over 8} (1+\ct)\Bigg)
+\Bigg[ {3 \over 8} -{3 \over 4} z +{3 \over 8} z^2
-{3\over 2} z \log(z) \Bigg]
\nonumber \\[2ex]
&&+ \Bigg[{3 \over 8} +{45 \over 4} z 
-{93 \over 8} z^2
+{21 \over 2} z \log(z) \Bigg]
\ct \Bigg\} \, ,
\nonumber \\[2ex]
\Delta^{(1)G \gamma/Z}_{q \bar q}&=&
\left({ \pi \over 8 N}\right) C_F 
\Bigg[ 
\Big(-12-12 z +{8 \over 1-z}\Big) \log(z)
+\Big(8 + 8 z\Big) \log(1-z)
\nonumber\\[2ex]&&
-16 \left( {\log(1-z) \over 1-z}\right)_+
+\Big(4+4 z -{8 \over (1-z)_+}-6 \delta(1-z)\Big) 
\log\left({Q^2 \over \mu_F^2}\right)
\nonumber\\[2ex]&&
-8 \zeta(2) \delta(1-z)
-12 +12 z +18 \delta(1-z) 
\Bigg]\Bigg(1 -3 \ct\Bigg) \, ,
\nonumber\\[2ex]
\Delta^{(1)G\gamma/Z}_{ q (\bar q)g}&=&
\left({\pi \over 8 N}\right)
T_F \Bigg[ 
\Big(-6+4 z^2\Big) \log(z)
+\Big(-4+8 z -8 z^2\Big) \log(1-z)
\nonumber\\[2ex]&&
+\Big(-2+4 z -4 z^2\Big) \log\left({Q^2 \over \mu_F^2}\right)
-5-2 z +7 z^2
\Bigg]\Bigg(1 -3 \ct\Bigg) \, ,
\nonumber\\[2ex]
\Delta^{(1),G\gamma/Z}_{gq(\bar q)}&=&
\left({\pi \over 8 N}\right)
T_F\Bigg[
\Big(2-28 z+4 z^2\Big) \log(z)
+\Big(-4+8 z-8 z^2\Big) \log(1-z)
\nonumber\\[2ex]&&
+\Big(-2+4 z -4 z^2\Big) \log\left({Q^2 \over \mu_F^2}\right)
-9-22 z +31 z^2 
\Bigg]\Bigg(1-3 \ct\Bigg) \, ,
\nonumber\\[2ex]
\Delta^{(1)G}_{q \bar q}&=&
\left({\pi \over 8 N}\right)  C_F\Bigg\{
\Bigg[ 
\Big({5 \over 2}+{5 \over 2} z -{5 \over 1-z}\Big)\log(z)
+\Big(-5-5 z\Big) \log(1-z)
\nonumber\\[2ex]&&
+10 \left({\log(1-z)\over 1-z}\right)_+
+\Big(-{5 \over 2} -{5 \over 2} z + {5 \over (1-z)_+}
   +{15 \over 4}\delta(1-z)\Big)
\nonumber\\[2ex]&&
\times \log\left({Q^2 \over \mu_F^2}\right)
+5 \zeta(2) \delta(1-z)
-{15 \over 2} +{10 \over 3 z} +{15 \over 2} z 
    -{10 \over 3} z^2 
\nonumber\\[2ex]&&
-{25 \over 2}\delta(1-z)\Bigg]
+\Bigg[
\Bigg({45 \over 2} +{45 \over 2} z +{15 \over 1-z} \Bigg)\log(z)
\nonumber\\[2ex]&&
+\Big(15 +15 z\Big) \log(1-z)
-30 \left({\log(1-z)\over 1-z}\right)_+
+\Bigg({15 \over 2} 
+{15 \over 2} z 
\nonumber\\[2ex]&&
-{15 \over (1-z)_+} 
-{45 \over 4} \delta(1-z) \Bigg) 
\log\left({Q^2 \over \mu_F^2}\right)
-15 \zeta(2) \delta(1-z)
+{225 \over2 }
\nonumber\\[2ex]&&
-{225 \over 2} z 
+{75 \over 2} \delta(1-z)
\Bigg]\ct\
+\Bigg[
\Big(-40 -40 z -{20 \over 1-z}\Big) \log(z)
\nonumber\\[2ex]&&
+\Big(-20-20 z\Big) \log(1-z)
+40 \left({\log(1-z)\over 1-z}\right)_+
+\Big(-10-10 z 
\nonumber\\[2ex]&&
+{20 \over (1-z)_+} +15 \delta(1-z)\Big)
\log\left({Q^2 \over \mu_F^2}\right)
+20 \zeta(2) \delta(1-z)
-150
\nonumber\\[2ex]&&
-{10 \over 3 z}
+150 z +{10 \over 3} z^2 -50 \delta(1-z) 
\Bigg]\cf \Bigg\}\, ,
\nonumber \\[2ex]
\Delta^{(1)G}_{q(\bar q) g}&=&
{\pi \over 8 N} T_F  \Bigg\{
\Bigg[
\Big({35 \over 4} -{10 \over z} -10 z -{5 \over 2} z^2\Big)\log(z)
+\Big(-{35 \over 2} +{20 \over z}+5 z+5 z^2\Big) 
\nonumber \\[2ex]&&
\times \log(1-z)
+\Big(-{35 \over 4} +{10 \over z}+{5 \over 2} z+{5 \over 2} z^2\Big)
\log\left({Q^2 \over \mu_F^2}\right)
\nonumber \\[2ex]&&
-{15 \over 8}-{15 \over 2 z} +{75 \over 4} z -{35 \over 8} z^2
\Bigg]
+\Bigg[
\Big({135 \over 4} +45 z +{15 \over 2} z^2\Big)\log(z)
\nonumber \\[2ex]&&
+\Big(-{15 \over 2} +15 z-15 z^2\Big) \log(1-z)
+\Big(-{15 \over 4} +{15 \over 2} z-{15 \over 2} z^2\Big)
\log\left({Q^2 \over \mu_F^2}\right)
\nonumber \\[2ex]&&
+{645 \over 8}-{375 \over 4} z +{105 \over 8} z^2
\Bigg]\ct
+\Bigg[
\Big(-65 +{10 \over z} -35 z -10 z^2\Big)\log(z)
\nonumber \\[2ex]&&
+\Big(30 -{20 \over z}-30 z+20 z^2\Big) 
\log(1-z)
+\Big(15 -{10 \over z}-15 z+10 z^2\Big)
\nonumber \\[2ex]&&
\times \log\left({Q^2 \over \mu_F^2}\right)
-{205 \over 2}+{15 \over 2 z} +{215 \over 2} z -{35 \over 2} z^2
\Bigg]\cf \Bigg\} \, ,
\nonumber \\[2ex]
\Delta^{(1)G}_{gq(\bar q)}&=&
{\pi \over 8 N} T_F \Bigg\{
\Bigg[
\Big({35 \over 4} -{10 \over z} +{25 \over 2} z -{5 \over 2} z^2\Big)\log(z)
+\Big(-{35 \over 2} +{20 \over z}+5 z+5 z^2\Big) 
\nonumber \\[2ex]&&
\times \log(1-z)
+\Big(-{35 \over 4} +{10 \over z}+{5 \over 2} z+{5 \over 2} z^2\Big)
\log\left({Q^2 \over \mu_F^2}\right)
\nonumber \\[2ex]&&
+{285 \over 8}-{35 \over 2 z} -{15 \over 4} z -{75 \over 8} z^2
\Bigg]
+\Bigg[
\Big(-{465 \over 4} -{345 \over 2} z +{15 \over 2} z^2\Big)\log(z)
\nonumber \\[2ex]&&
+\Big(-{15 \over 2} +15 z-15 z^2\Big) \log(1-z)
+\Big(-{15 \over 4} +{15 \over 2} z-{15 \over 2} z^2\Big)
\log\left({Q^2 \over \mu_F^2}\right)
\nonumber \\[2ex]&&
-{1695 \over 8}-{20 \over z}+{615 \over 4} z +{625 \over 8} z^2
\Bigg]\ct
+\Bigg[
\Big(185 +{10 \over z} +215 z -10 z^2\Big)
\nonumber \\[2ex]&&
\times \log(z)
+\Big(30 -{20 \over z}-30 z+20 z^2\Big) 
\log(1-z)
+\Big(15 -{10 \over z}
\nonumber \\[2ex]&&
-15 z+10 z^2\Big)
\log\left({Q^2 \over \mu_F^2}\right)
+{395\over 2}+{545 \over 6 z} -{385 \over 2} z -{605 \over 6} z^2
\Bigg]\cf \Bigg\} \, ,
\nonumber \\[2ex]
\Delta^{(1)G}_{g g}&=&\left({\pi \over 2 (N^2-1)}\right) 
\Bigg\{C_A \Bigg[ 
\Big(10 -{5 \over z}-5 z +5 z^2 -{5 \over 1-z} \Big)\log(z)
\nonumber \\[2ex]&&
+\Big(-20+{10 \over z}+10 z-10 z^2\Big) \log(1-z)
+10 \left({\log(1-z)\over 1-z}\right)_+
\nonumber \\[2ex]&&
+\Big(-10+{5 \over z}+5 z -5 z^2 +{5 \over (1-z)_+} 
    +{55 \over 12}\delta(1-z)\Big) \log\left({Q^2 \over \mu_F^2}\right)
\nonumber \\[2ex]&&
+5 \zeta_2 \delta(1-z)
+{25 \over 4} -{85 \over 12 z} -{25 \over 4} z+{85 \over 12} z^2
      -{1015 \over 72} \delta(1-z) 
\Bigg]
\nonumber \\[2ex]&&
+C_A \Bigg[ 
\Big(-30-30 z\Big)\log(z)
-60-{5 \over z} +60 z +5 z^2 
\Bigg]\ct
\nonumber \\[2ex]&&
+C_A \Bigg[
\Big(40+{5 \over z} +55 z -5 z^2 +{5 \over 1-z} \Big)\log(z)
+\Big(20-{10 \over z}
\nonumber \\[2ex]&&
-10 z+10 z^2\Big) \log(1-z)
-10 \left({\log(1-z)\over 1-z}\right)_+
+\Big(10-{5 \over z}-5 z 
\nonumber \\[2ex]&&
+5 z^2 -{5 \over (1-z)_+}
    -{55 \over 12}\delta(1-z)\Big) \log\left({Q^2 \over \mu_F^2}\right)
-5 \zeta(2) \delta(1-z)
\nonumber \\[2ex]&&
+{255 \over 4}+{305\over 12 z}-{255 \over 4} z -{305 \over 12} z^2
+{1015 \over 72} \delta(1-z)
\Bigg]\cf
\nonumber \\[2ex]&&
+T_F n_f \Bigg[
-{5 \over 3} \delta(1-z) \log\left({Q^2 \over \mu_F^2}\right)
+{175 \over 36}\delta(1-z)
\Bigg]
\nonumber \\[2ex]&&
+T_F n_f \Bigg[
{5 \over 3} \delta(1-z) \log\left({Q^2 \over \mu_F^2}\right)
-{175 \over 36}\delta(1-z)
\Bigg]\cf
\Bigg\} \, ,
\label{eq2.24}
\end{eqnarray}
where $C_F={(N^2 -1) / 2 N}$, $C_A=N$ and $T_F=1/2$ are the $SU(N)$ colour 
factors and $n_f$ is the number of flavours.  The "plus" functions appearing 
in the above results are the distributions which satisfy the following 
equation
\begin{eqnarray}
\int_0^1 dz~f_+(z)~ g(z) &=& \int_0^1 dz~f(z) \Big(g(z)-g(1)\Big) \, ,
\nonumber
\end{eqnarray}
where 
\begin{eqnarray}
f_+(z)&=&\left (\frac{\ln^i(1-z)}{1-z}\right )_+ , \quad \quad \quad  i=0,1\, 
\nonumber
\end{eqnarray}
and $g(z)$ is any well behaved function in the region $0\le z \le 1$.

In Eqs.~(\ref{eq2.23}) and (\ref{eq2.24}) the term $(1-3 \cos^2 \theta^*)$ 
corresponds to the interference between the SM and GR and within the SM 
interference between $\gamma$ and $Z$ diagrams.  Though this 
combination is even in $\cos \theta^*$ it vanishes in the angular 
integrated cross section and also does not contribute to the 
forward-backward asymmetry $A_{FB}$.  Hence the un-integrated cross 
section is very useful to study this contribution to the interference 
effect in the Drell-Yan process.  

We present below the angular distribution which is "odd" in $\co$:
\begin{eqnarray}
\label{eq2.27}
2 S~{d \sigma_{o}^{P_1P_2} \over dQ^2 d\cos \theta^*}&=&
\sum_q{\delta \cal F}_{SM,q} \int_0^1~ {d x_1}~ \int_0^1
~{dx_2}~ \int_0^1~ dz~ \delta(\tau-z x_1 x_2)
\nonumber\\[2ex]&&
\times \Bigg[ \delta H_{q \bar q}(x_1,x_2,\mu_F^2) \Big(
\delta \Delta_{q \bar q}^{(0),\gamma Z}(z,Q^2,\mu_F^2)
 +a_s \delta \Delta_{q \bar q}^{(1),\gamma Z}(z,Q^2,\mu_F^2)\Big)
\nonumber\\[2ex] &&
+\delta H_{q g}(x_1,x_2,\mu_F^2)
 \Big(a_s \delta \Delta_{q g}^{(1),\gamma Z}(z,\mu_F^2) \Big)
\nonumber\\[2ex] &&
+\delta H_{g q}(x_1,x_2,\mu_F^2)
 \Big(a_s \delta \Delta_{g q}^{(1),\gamma Z}(z,\mu_F^2) \Big)
\Bigg]
\nonumber\\[2ex]
&&+\sum_q{\delta \cal F}_{SMGR,q} \int_0^1~ {d x_1 }~ \int_0^1
~{dx_2}~ \int_0^1~ dz~ \delta(\tau-z x_1 x_2)
\nonumber\\[2ex]&&
\times \Bigg[ \delta H_{q \bar q}(x_1,x_2,\mu_F^2) \Big(
\delta \Delta_{q \bar q}^{(0),G \gamma/Z}(z,Q^2,\mu_F^2)
 +a_s \delta \Delta_{q \bar q}^{(1),G \gamma/Z}(z,Q^2,\mu_F^2)\Big)
\nonumber\\[2ex] &&
+\delta H_{q g}(x_1,x_2,\mu_F^2)
 \Big(a_s\delta \Delta_{q g}^{(1),G \gamma/Z}(z,\mu_F^2) \Big)
\nonumber\\[2ex] &&
+\delta H_{g q}(x_1,x_2,\mu_F^2)\Big)
 \Big(a_s\delta \Delta_{g q}^{(1),G \gamma/Z}(z,\mu_F^2)\Big)
 \Bigg]\,.
\end{eqnarray}
The constants $\delta {\cal F}_{SM,q},\delta {\cal F}_{SMGR,q}$ are given by
\begin{eqnarray}
\delta {\cal F}_{SM,q}&=&{2 \alpha^2 } \Bigg[{(Q^2-M_Z^2) \over
\left((Q^2-M_Z^2)^2+M_Z^2 \Gamma_Z^2\right)c_w^2 s_w^2} Q_q Q_e g_q^A g_e^A
\nonumber\\[2ex]
&&+{2 Q^2 \over
\left((Q^2-M_Z^2)^2+M_Z^2 \Gamma_Z^2\right) c_w^4 s_w^4}
g_q^V g_e^V g_q^A g_e^A
\Bigg]\,,
\\[2ex]
\delta {\cal F}_{SMGR,q}&=&{\alpha \kappa^2 Q^2 \over 4 \pi }|{\cal D}(Q^2)|
\Bigg[Q_q Q_e +
{Q^2 (Q^2-M_Z^2) \over
\left((Q^2-M_Z^2)^2+M_Z^2 \Gamma_Z^2\right)c_w^2 s_w^2}
g_q^V g_e^V \Bigg] \,.
\end{eqnarray}
The renormalised incoming partonic fluxes are defined by
\begin{eqnarray}
\label{eq2.30}
\delta H_{q \bar q}(x_1,x_2,\mu_F^2)&=&
f_q^{P_1}(x_1,\mu_F^2)~
f_{\bar q}^{P_2}(x_2,\mu_F^2)
-f_{\bar q}^{P_1}(x_1,\mu_F^2)~
f_q^{P_2}(x_2,\mu_F^2)\,,
\nonumber
\\[2ex]
\delta H_{g q}(x_1,x_2,\mu_F^2)&=&
f_g^{P_1}(x_1,\mu_F^2) ~
\Big(f_q^{P_2}(x_2,\mu_F^2)
-f_{\bar q}^{P_2}(x_2,\mu_F^2)\Big)\,,
\nonumber
\\[2ex]
\delta H_{q g}(x_1,x_2,\mu_F^2)&=&
\delta H_{g q}(x_2,x_1,\mu_F^2)\,.
\end{eqnarray}
The LO coefficient functions corresponding to Eq.~(\ref{eq2.27}) are 
\begin{eqnarray}
\label{eq2.31}
\delta \Delta^{(0),\gamma Z}_{q\bar q}&=&{2 \pi \over N} \delta(1-z)
\Big[\co\Big]\,,
\nonumber \\[2ex]
\delta \Delta^{(0),G \gamma/Z}_{q \bar q}&=&{\pi \over 8 N} \delta(1-z) 
\Big[2 \cth\Big]\,.
\end{eqnarray}
The NLO contributions are given by
\begin{eqnarray}
\label{eq2.32}
\delta \Delta^{(1)\gamma Z}_{q \bar q}&=&
{2 \pi \over N} C_F \Bigg[
\Big(8+ 8 z -{8 \over 1-z}\Big) \log(z)
+\Big(-8-8 z\Big) \log(1-z)
\nonumber \\[2ex]&&
+16 \left( {\log(1-z) \over 1-z}\right)_+
+\Big(-4 -4 z +{8 \over(1-z)_+ }+6 \delta(1-z))
\log\left({Q^2 \over \mu_F^2}\right)
\nonumber \\[2ex]&&
+8 \zeta(2) \delta(1-z)
+4 -4 z -16 \delta(1-z)
\Bigg]\co \,,
\nonumber\\[2ex]
\delta \Delta^{(1)\gamma Z}_{q g}&=&
{2 \pi \over N} T_F
\Bigg[
\Big(2-4 z^2\Big) \log(z)
+\Big(4-8 z +8 z^2\Big) \log(1-z)
\nonumber \\[2ex]&&
+\Big(2-4 z +4 z^2\Big) \log\left({Q^2 \over \mu_F^2}\right)
+1 +6 z -7 z^2
\Bigg]\co \,,
\nonumber\\[2ex]
\delta \Delta^{(1)\gamma Z}_{g q}&=&
{2 \pi \over N} T_F
\Bigg[
\Big(2-4 z + 12 z^2\Big) \log(z)
+\Big(-4+8 z-8 z^2\Big) \log(1-z)
\nonumber \\[2ex]&&
+\Big(-2+4 z -4 z^2\Big) \log\left({Q^2 \over \mu_F^2}\right)
+1 -2 z+ z^2
\Bigg]\co \,,
\nonumber\\[2ex]
\delta \Delta^{(1) G \gamma/Z}_{q \bar q}&=&
{\pi \over 8 N} C_F \Bigg\{
\Bigg[- {16 \over 1-z} \log(z) 
+\Big(-16-16 z\Big) \log(1-z)
\nonumber \\[2ex]&&
+32\left( {\log(1-z) \over 1-z}\right)_+
+\Big(-8 -8 z +{16 \over(1-z)_+ }+12 \delta(1-z))
\log\left({Q^2 \over \mu_F^2}\right)
\nonumber \\[2ex]&&
+16 \zeta(2) \delta(1-z)
-48 +48 z -36 \delta(1-z) 
\Bigg] \cth
\nonumber \\[2ex]&&
+ \Bigg[24 -24 z \Bigg]\co\Bigg\} \,,
\nonumber\\[2ex]
\delta \Delta^{(1) G \gamma/Z}_{q g}&=&
{\pi \over 8 N} T_F \Bigg\{
\Bigg[
\Big(-12-24 z -8 z^2\Big) \log(z)
+\Big(8-16 z+16 z^2\Big) \log(1-z)
\nonumber \\[2ex]&&
+\Big(4-8 z +8 z^2\Big) \log\left({Q^2 \over \mu_F^2}\right)
-38 +52 z -14 z^2
\Bigg] \cth
\nonumber \\[2ex]&&
+ \Bigg[ 24 z \log(z) +24 -24 z \Bigg] \co \Bigg\} \,,
\nonumber\\[2ex]
\delta \Delta^{(1) G \gamma/Z}_{g q}&=&
{\pi \over 8 N} T_F \Bigg\{
\Bigg[
\Big(36 +72 z +24 z^2 \Big) \log(z)
+\Big(-8+16 z-16 z^2\Big) \log(1-z)
\nonumber \\[2ex]&&
+\Big(-4+8 z-8 z^2\Big) \log\left({Q^2 \over \mu_F^2}\right)
+98-100 z +2 z^2
\Bigg]\cth
\nonumber \\[2ex]&&
+ \Bigg[ -48 z \log(z) -48 +48 z \Bigg]\co \Bigg\} \,.
\end{eqnarray}
These coefficient functions Eq.~(\ref{eq2.31}) and (\ref{eq2.32}), which 
are odd in $\cos \theta^*$, are due to the interference of $\gamma$ and $Z$ 
in SM and between $SM$ and $GR$ in the full theory.  Note that the $q \to
\bar q$ in the $qg$ subprocess leads to a negative sign which has been taken
care of in the renormalised parton fluxes Eq.~(\ref{eq2.30}).  $A_{FB}$ picks 
up this odd parts which contributes to the interference terms.  
To NLO the $A_{FB}$ coefficient functions have been evaluated in \cite{us1} 
\footnote{In \cite{us1} the last three equations of Eq.~(6.11), the RHS 
should read $\Delta_{ab}^{(1)\gamma/Z}$.}
and the effects analysed for the ADD case.  
In the next section, the impact of the NLO-QCD correction derived in 
this section is discussed.

\mysection{Discussions}

In this section, the effect of the NLO QCD corrections on the angular 
distribution of lepton pair are presented. We present these distributions 
for the LHC ($\sqrt{S}=14~{\rm TeV}$) and Run II of Tevatron ($\sqrt{S}=
1.96~{\rm TeV}$) for typical values of ADD and RS model parameters.  The 
effort here is mainly to emphasise the impact of QCD correction on the 
bounds rather than to extract bounds on $M_S$.

For ADD model, we choose the parameters $M_S=2$ TeV and $d=3$.  For RS we 
choose $c_0=0.01$, $M_1=1.5$ TeV(for LHC) and $M_1=700$ GeV(for Tevatron).  
The SM parameters which enter our analysis are $\alpha= 1/137.03604$, 
$M_Z=91.1876$ GeV, $\Gamma_Z=2.4952$ GeV and $\sin^2\theta_W=0.227$.  
For the parton density sets, we adopt in leading order, the MRST 2001 LO 
($\Lambda=0.1670$ GeV) and in next-to-leading order, the MRST 2001 NLO 
($\Lambda=0.2390$ GeV).  The renormalisation scale $\mu_R$ and factorisation 
scale $\mu_F$ are taken to be equal to $Q$ unless mentioned otherwise.

For the coefficient functions which are even in $\cos \theta^*$, the parton 
density combinations are even under the interchange of $x_1$ and $x_2$, 
while for the odd terms in $\cos \theta^*$, the parton density 
combinations are odd under this exchange.  Hence, the quark-antiquark 
initiated contributions from these odd terms to LHC cross sections are zero,
but small contribution comes from quark-gluon initiated processes.  This is not 
the case for Tevatron.

In the SM part, at LO level, the quark-antiquark initiated processes behave 
as $1+\cos^2 \theta^*$ for pure $\gamma$ and $Z$ intermediate states, and as 
$\cos \theta^*$ for $\gamma Z$ interference terms.  In the Gravity part, at LO,
gluon-gluon initiated process is of the form $1-\cos^4 \theta^*$, 
quark-antiquark process is of the form $1-3 \cos^2 \theta^*+4 \cos^4 \theta^*$.
The interference between SM and GR always behaves as $\cos \theta^*$, $\cos^3
\theta^*$ and $1-3 \cos^2 \theta^*$.  At NLO, quark-gluon initiated 
processes contribute to both SM, GR and the interference terms.  

We first discuss the phenomenology at LHC using ADD model.  In Fig.~1a, we plot 
the angular distribution at $Q=700$ GeV with NLO corrected cross sections.  We 
find that the gravity contribution is large compared to that of SM.   Since 
the gluon flux is large at LHC, the gluon-gluon initiated subprocess dominates 
over the rest.  Also, the interference between SM and Gravity is negligible
over the entire range of $\cos \theta^*$ at this $Q=700$ GeV.   
In the Fig.~1b, we have plotted the K-factor at LHC.  In general K-factor is 
defined as
\begin{eqnarray}
K= \left [ {d \sigma_{LO}^I (Q,\cos \theta^*) 
\over d Q d\cos \theta^*}\right]^{-1}
\left [ {d \sigma_{NLO}^I (Q,\cos \theta^*) 
\over d Q d\cos \theta^*}\right] ~,
\label{kfact}
\end{eqnarray}
where $I=SM,TOT$,
SM means Standard Model, TOT means sum of SM and Gravity contributions.  
Since the gluon-gluon initiated process dominates over the rest the 
K-factor is around $1.4$ to $1.5$ in the entire range of $\cos \theta^*$.
In the Fig.~1c we have plotted the R-ratio in order to check whether the 
NLO results improve the scale uncertainty.  Here we have chosen $\mu_R=
\mu_F=\mu$ and we follow the same throughout our analysis.  The R-ratio 
is defined as
\begin{eqnarray}
R^I_{LO}&=& \left [ {d \sigma_{LO}^I (\mu=\mu_0))
\over d Q d\cos \theta^*}\right]^{-1}
\left [ {d \sigma_{LO}^I (\mu))
\over d Q d\cos \theta^*}\right]_{Q=700~GeV} ~,
\label{rfact1}
\\[2ex]
R_{NLO}&=& \left [ {d \sigma_{NLO}^I (\mu=\mu_0))
\over d Q d\cos \theta^*}\right]^{-1}
\left [ {d \sigma_{NLO}^I (\mu))
\over d Q d\cos \theta^*}\right]_{Q=700~GeV} ~.
\label{rfact2}
\end{eqnarray}
We have chosen $\theta^*=45^0$ for the plot.
As we can see the NLO results improve the scale uncertainty.

Let us now repeat the similar study for the RS model at LHC energies.
In the Fig.~2a, we have plotted the angular distribution at first resonance 
$M_1=1.5$ TeV and $c_0=0.01$.  We find that the gravity contribution is well 
above the standard model one.  In particular, the gluon-gluon initiated 
contribution is the dominant one.  Since the SM and interference 
contributions are of the same order and are extremely small due to large $Q$ 
which is $1.5$ TeV, the total contribution is purely due to the gluon-gluon 
initiated process.  Unlike the ADD case (see Fig.~1a), the total contribution 
mainly comes from the gravity mediated process at the first resonance.
In Fig.~2b, we have plotted the K-factor defined in Eq.~(\ref{kfact}) at
the first resonance.  Because of this, the dominant contribution comes from the 
gravity mediated processes.  Since $Q=1.5$ TeV, both quark-antiquark as 
well as the gluon initiated processes contribute at the same level because 
their partonic fluxes are comparable at this energy.  The shape of the 
K-factor in Fig.~1b looks different from Fig.~2b because at $Q=0.7$ TeV, only 
gluon initiated process dominates.  In Fig.~2c, we have plotted the R-ratio 
defined in Eq.~(\ref{rfact1}, \ref{rfact2}) for RS resonance $M_1=1.5$ TeV,
$c_0=0.01$ at $\theta^*=45^0$.  From the plot it is clear that the NLO 
corrections improve the scale uncertainty.    

Next we discuss the phenomenology at Tevatron.  We start with ADD for the 
parameters $M_S=2$ TeV, $d=3$ and $Q=400$ GeV.  In Fig.~3a we have plotted 
angular distribution with NLO results.  We find that the SM dominates over 
the rest.  In Tevatron, both the even and odd in $\cos \theta^*$ contribute 
significantly unlike in the LHC.  This leads to the asymmetry in the angular 
distribution as shown in Fig.~3a at Tevatron.  At $Q=400$ GeV the dominant 
contribution at Tevatron is from SM.  
In Fig.~3b, we have plotted the K-factor at $Q=400$ GeV.  Since it is the 
quark-antiquark initiated process that dominates, the K-factor is similar 
to the SM value which is around 1.3.  Fig.~3c shows the sensitivity of the 
results to the scale variation.  As expected NLO improves the result.

We now study the phenomenology at Tevatron in the case of RS model.  We 
choose the RS model parameters $M_1=700$ GeV and $c_0=0.01$ which are
not excluded by the recent searches by D\O \cite{d0_rs}.  In Fig.~4a, 
we have plotted the 
angular distribution at the first resonance using NLO corrected cross 
sections.  Being in the resonance region, gravity mediated process 
dominates over the SM.  At $Q=700$ GeV, the quark initiated processes 
contribute significantly, as the $q \bar q$ flux dominates.
In Fig.~4b, we have plotted the K-factor at $Q=700$ GeV.  Since the quark 
initiated process dominates, the K-factor is close to the SM value.
Fig.~4c shows the sensitivity of the results with respect to scale.  
One can easily notice that NLO gives reliable predictions.
 
In summary, we have computed the cross sections $d\sigma/dQ/d \cos \theta^*$
up to next to leading order in QCD. Along with the standard model results, we 
have presented the contributions from all the subprocesses that are due to the 
graviton in the context of TeV-scale gravity models.  Our main conclusion 
is that the NLO QCD corrections are very significant at the LHC because of 
the large incident gluon flux.  At the Tevatron where the gluon flux is small, 
the NLO effects are moderate for ADD and RS in the resonance region.  But, 
significantly, at both the colliders the inclusion of the NLO QCD corrections 
help stabilise the cross-section with respect to scale variations.  The 
extraction of bounds from both the colliders will, therefore, require the 
inclusion of these NLO QCD corrections.\\[3mm]

\noindent
{\it Acknowledgments}:\\
\noindent 
The work of PM is part of a project (IFCPAR Project No. 2904-2) on 
`Brane-World Phenomenology' supported by the Indo-French Centre for 
the Promotion of Advanced Research, New Delhi, India.  We thank 
S.~Raychaudhuri for providing the code that evaluates the RS KK mode 
sum in the propagator.

\eject
\centerline{\large \bf Figure Caption}
\vspace{.5cm}

\noindent
Figure 1. 
(a) The double differential cross section $d\sigma/dQ/d \cos \theta^*$ 
is plotted as a function of $\cos \theta^*$ for $Q=700$ GeV at LHC.  The 
typical ADD parameters chosen are $M_S=2$ TeV, $d=3$.  
(b) The corresponding K-factor for $\cos \theta^*$ distribution SM and 
SM plus gravity (TOT). 
(c) Scale variation at LO and NLO as defined in Eq.~(\ref{rfact1}), 
(\ref{rfact2}) for $Q=700$ GeV and $\theta^* =45^0$. \\[1ex]

\noindent
Figure 2. (a) The double differential cross section $d \sigma /dQ /d 
\cos \theta^*$ is plotted as a function of $\cos \theta^*$ for $Q=1.5$ 
TeV at LHC.  The RS model parameters are $M_1=1.5$ TeV and $c_0=0.01$.
(b) The K-factor for the distribution in (a) is plotted for the $\cos 
\theta^*$ range [-1,1].  (c) The scale variation of the ratio R is 
plotted as a function of $\mu/\mu_0$ at the first RS KK resonance 
region and $\theta^*=45^0$. \\[1ex]

\noindent
Figure 3. (a) For Tevatron energies, $d \sigma /dQ /d \cos \theta^*$ 
is plotted as a function of $\cos \theta^*$ at $Q=400$ GeV for typical
value of ADD parameters $M_S=2$ TeV and $d=3$. (b) The K-factor for 
$\cos \theta^*$ distribution for the same ADD parameters in (a) is 
plotted. (c) The variation of the R-ratio with respect to the 
scale  $\mu/\mu_0$ for the ADD parameters in (a) at $\theta^*=45^0$.
\\[1ex]

\noindent
Figure 4. (a) The double differential cross section $d \sigma /dQ /d \cos 
\theta^*$ is plotted as a function of $\cos \theta^*$ for $Q=700$ GeV at the 
Tevatron.  RS parameters $M_1=700$ GeV and $c_0=0.01$.  
(b) The K-factor for the distribution in (a) is plotted for the $\cos 
\theta^*$ range.  (c) The scale variation of the ratio R is plotted as a 
function of $\mu/\mu_0$ for $\theta^*=45^0$ and $Q=700$ GeV.

\eject





\begin{figure}[htb]
\vspace{1mm}
\centerline{\epsfig{file=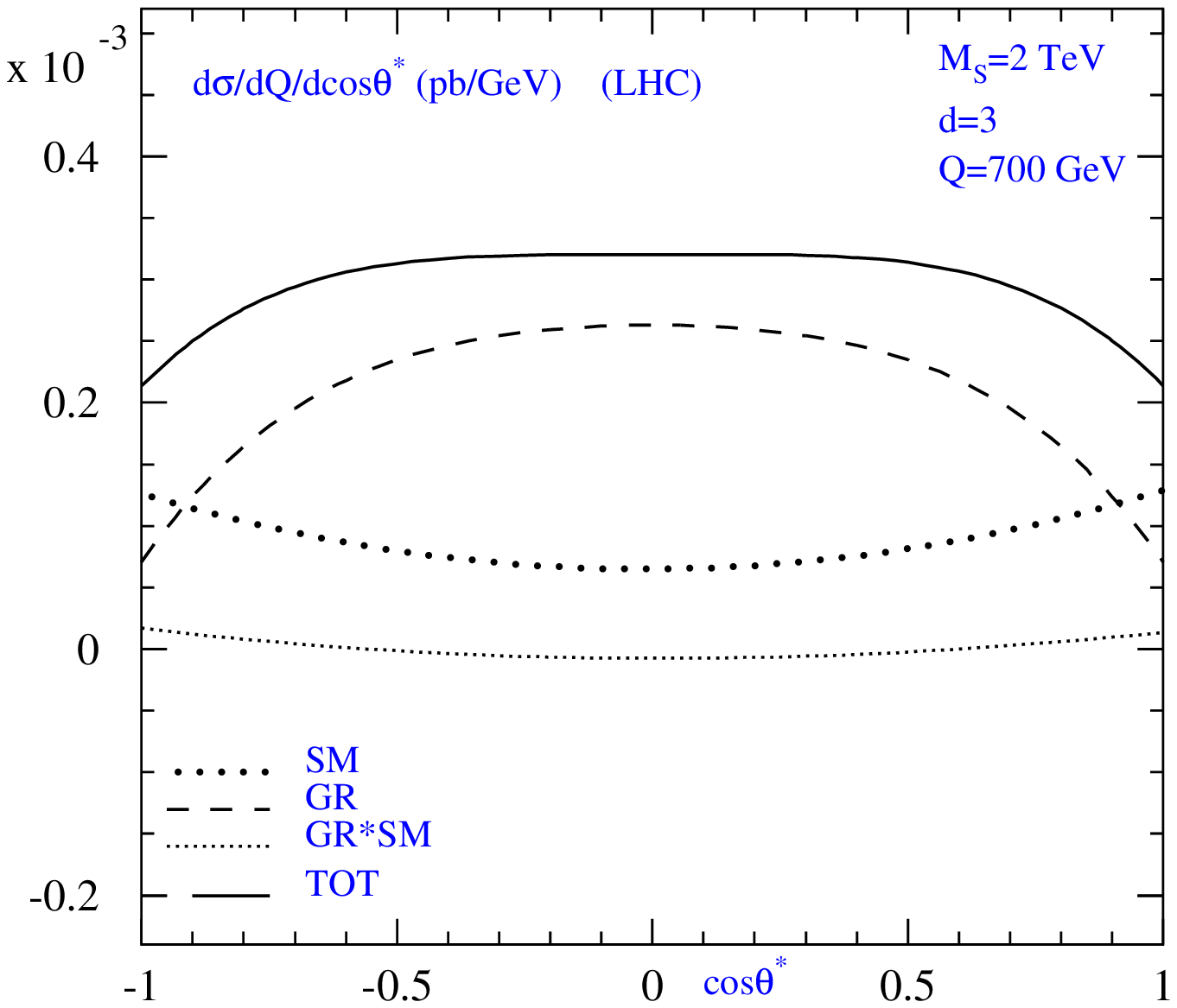,width=15cm,height=16cm,angle=0}}
\vspace{5mm}
\centerline{\bf Fig.~1a}
\end{figure}
                                                                                
\eject
                                                                                
\begin{figure}[htb]
\vspace{1mm}
\centerline{\epsfig{file=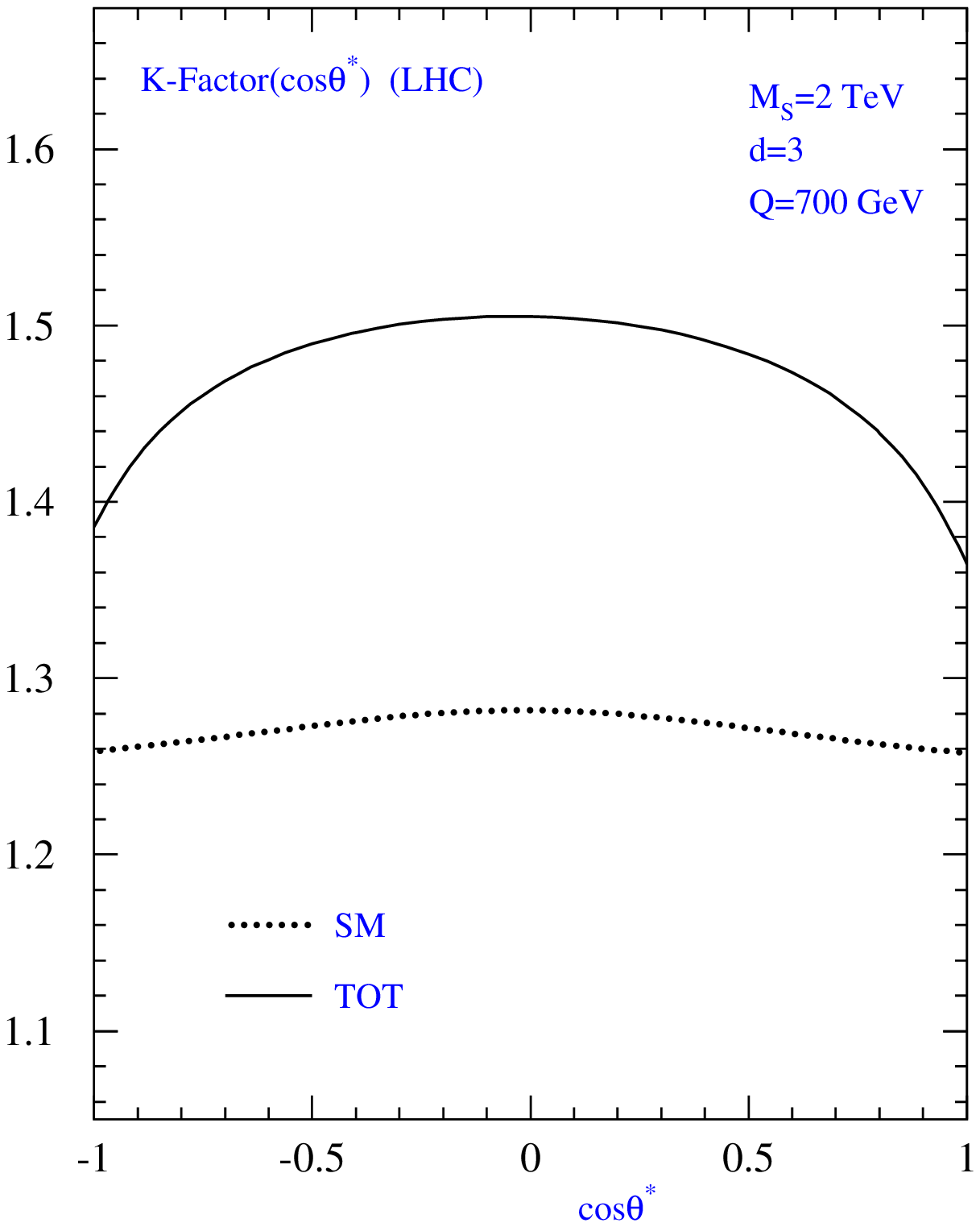,width=15cm,height=16cm,angle=0}}
\vspace{5mm}
\centerline{\bf Fig.~1b}
\end{figure}
                                                                                
\eject
\begin{figure}[htb]
\vspace{1mm}
\centerline{\epsfig{file=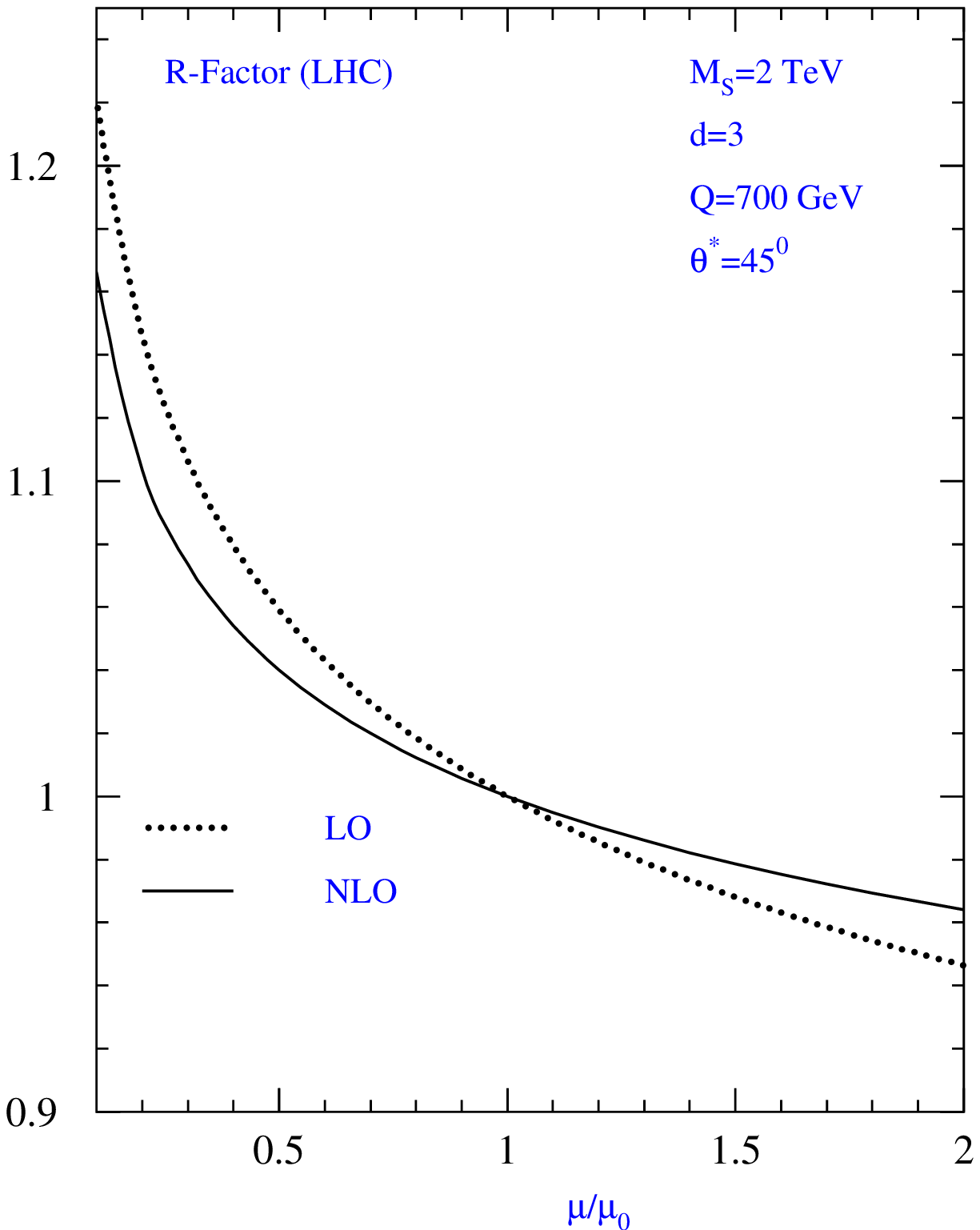,width=15cm,height=16cm,angle=0}}
\vspace{5mm}
\centerline{\bf Fig.~1c}
\end{figure}
                                                                                
\eject
\begin{figure}[htb]
\vspace{1mm}
\centerline{\epsfig{file=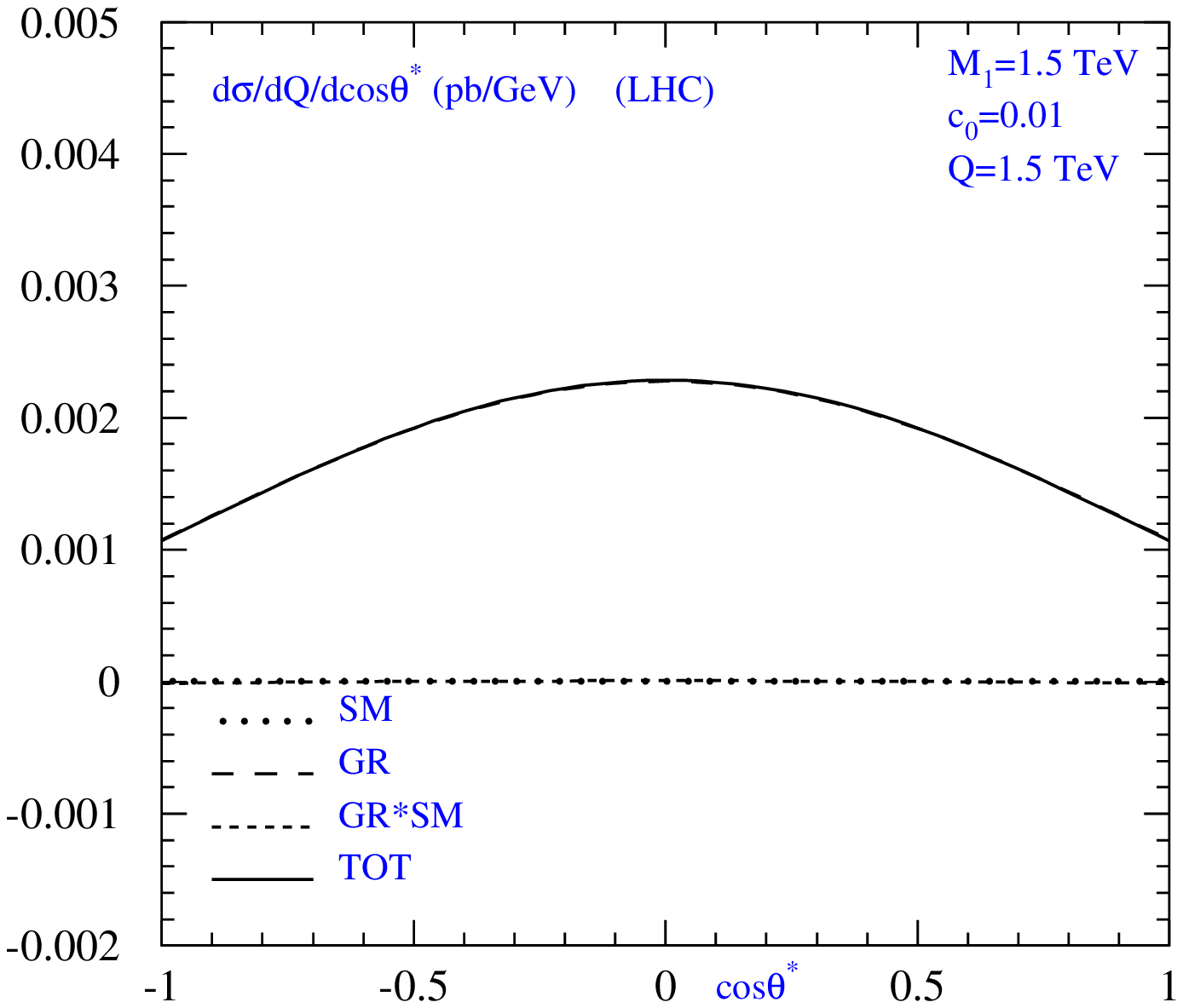,width=15cm,height=16cm,angle=0}}
\vspace{5mm}
\centerline{\bf Fig.~2a}
\end{figure}
                                                                                
\eject
\begin{figure}[htb]
\vspace{1mm}
\centerline{\epsfig{file=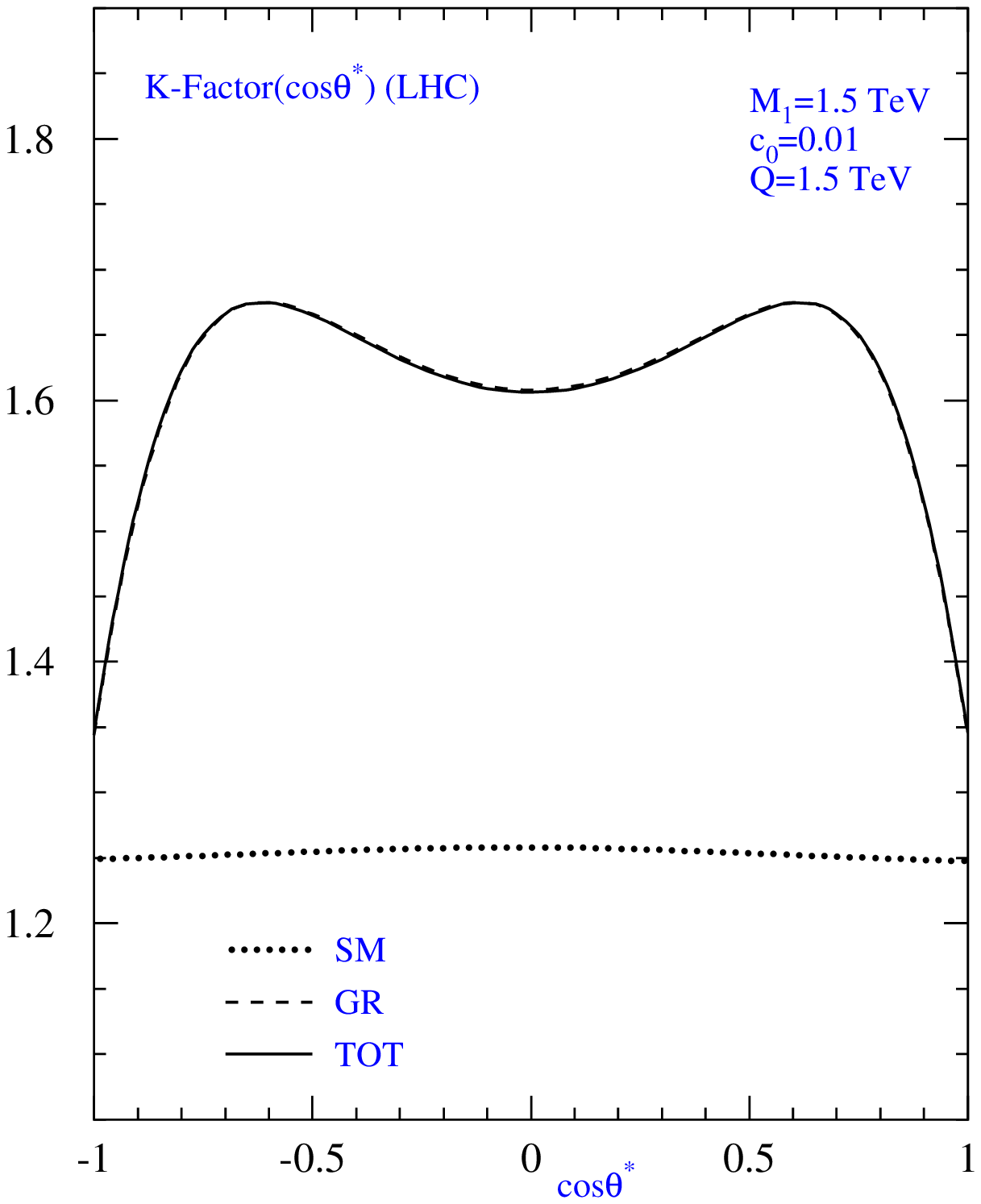,width=15cm,height=16cm,angle=0}}
\vspace{5mm}
\centerline{\bf Fig.~2b}
\end{figure}
                                                                                
\eject
\begin{figure}[htb]
\vspace{1mm}
\centerline{\epsfig{file=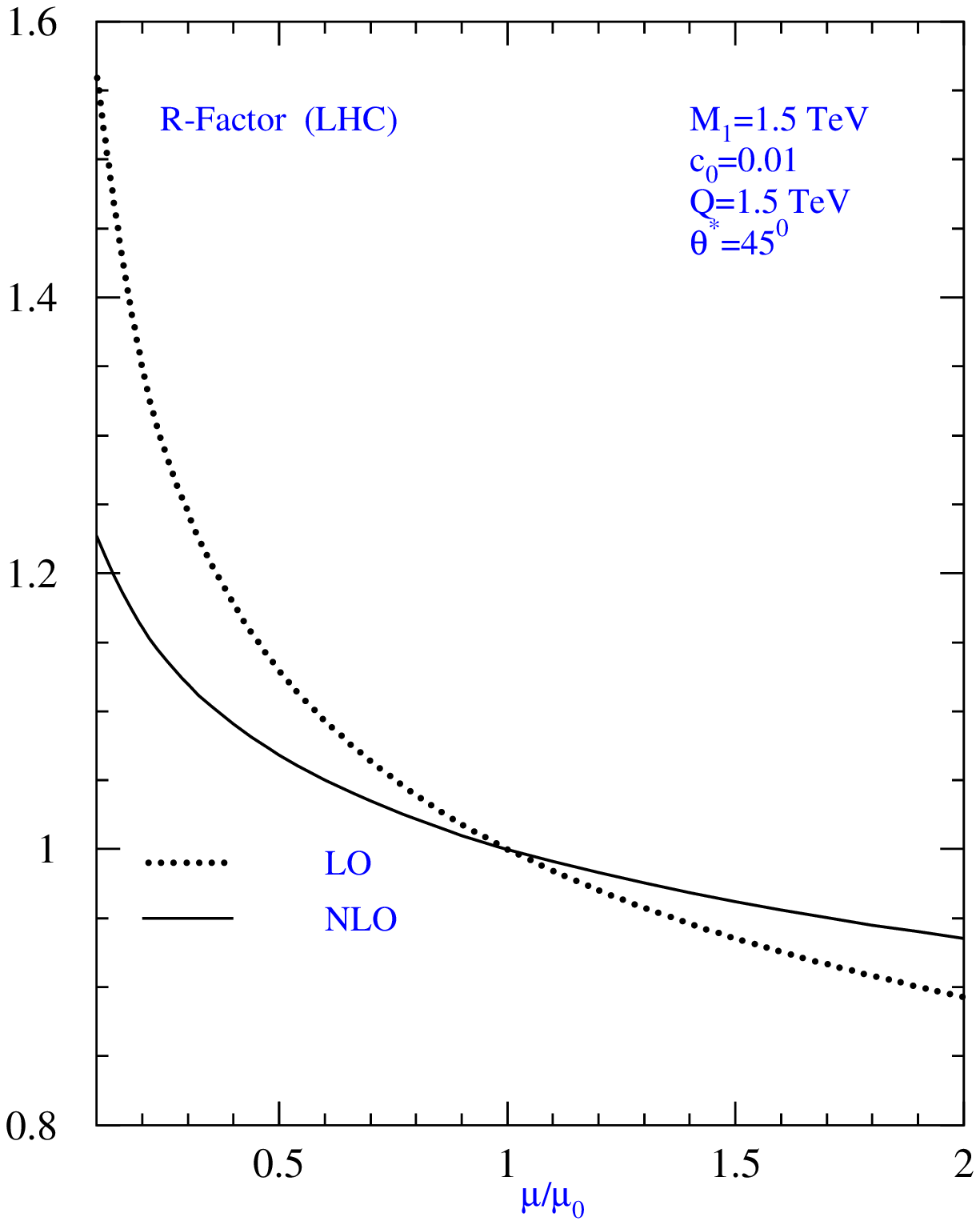,width=15cm,height=16cm,angle=0}}
\vspace{5mm}
\centerline{\bf Fig.~2c}
\end{figure}
                                                                                
\eject
\begin{figure}[htb]
\vspace{1mm}
\centerline{\epsfig{file=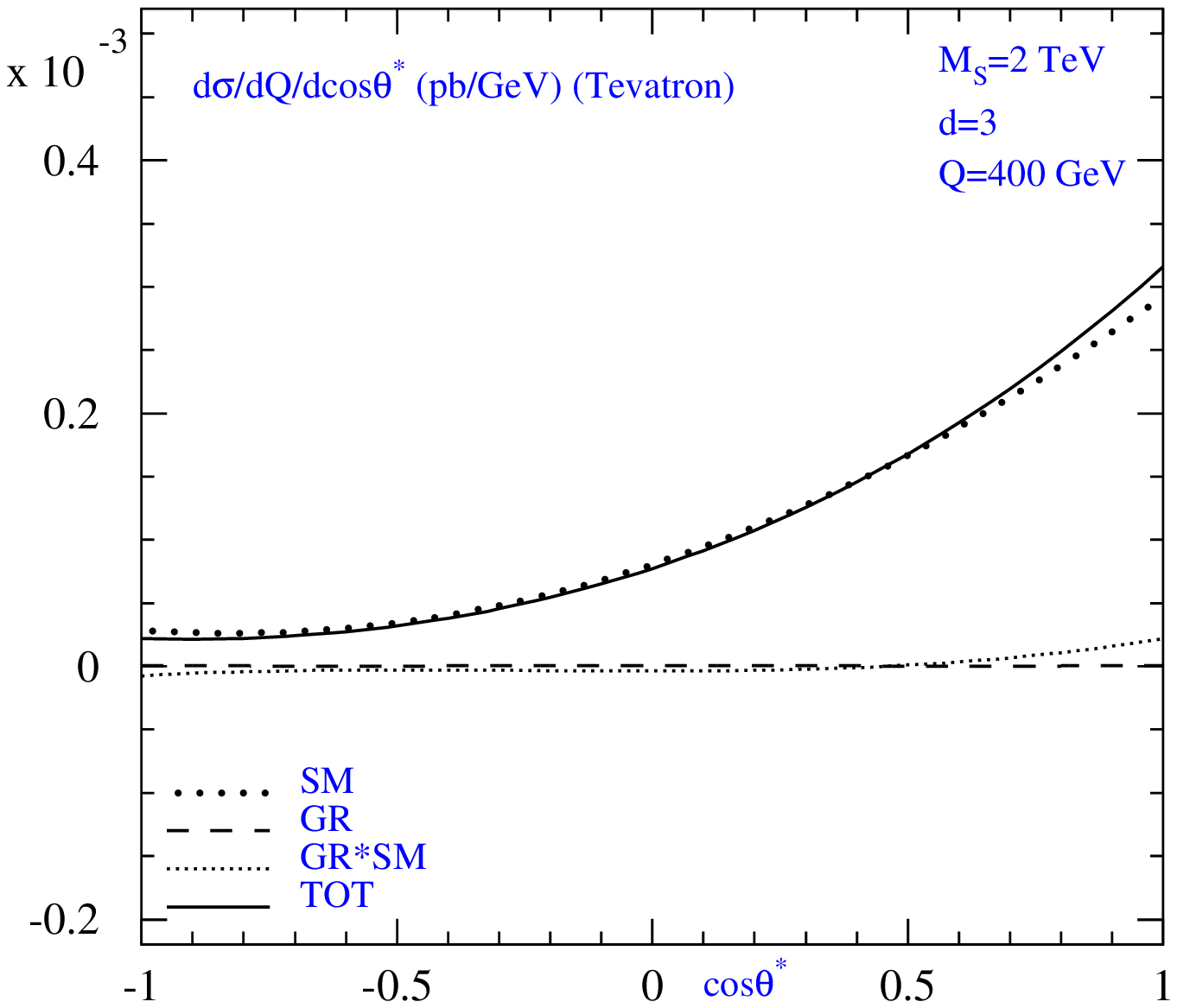,width=15cm,height=16cm,angle=0}}
\vspace{5mm}
\centerline{\bf Fig.~3a}
\end{figure}
                                                                                
\eject
\begin{figure}[htb]
\vspace{1mm}
\centerline{\epsfig{file=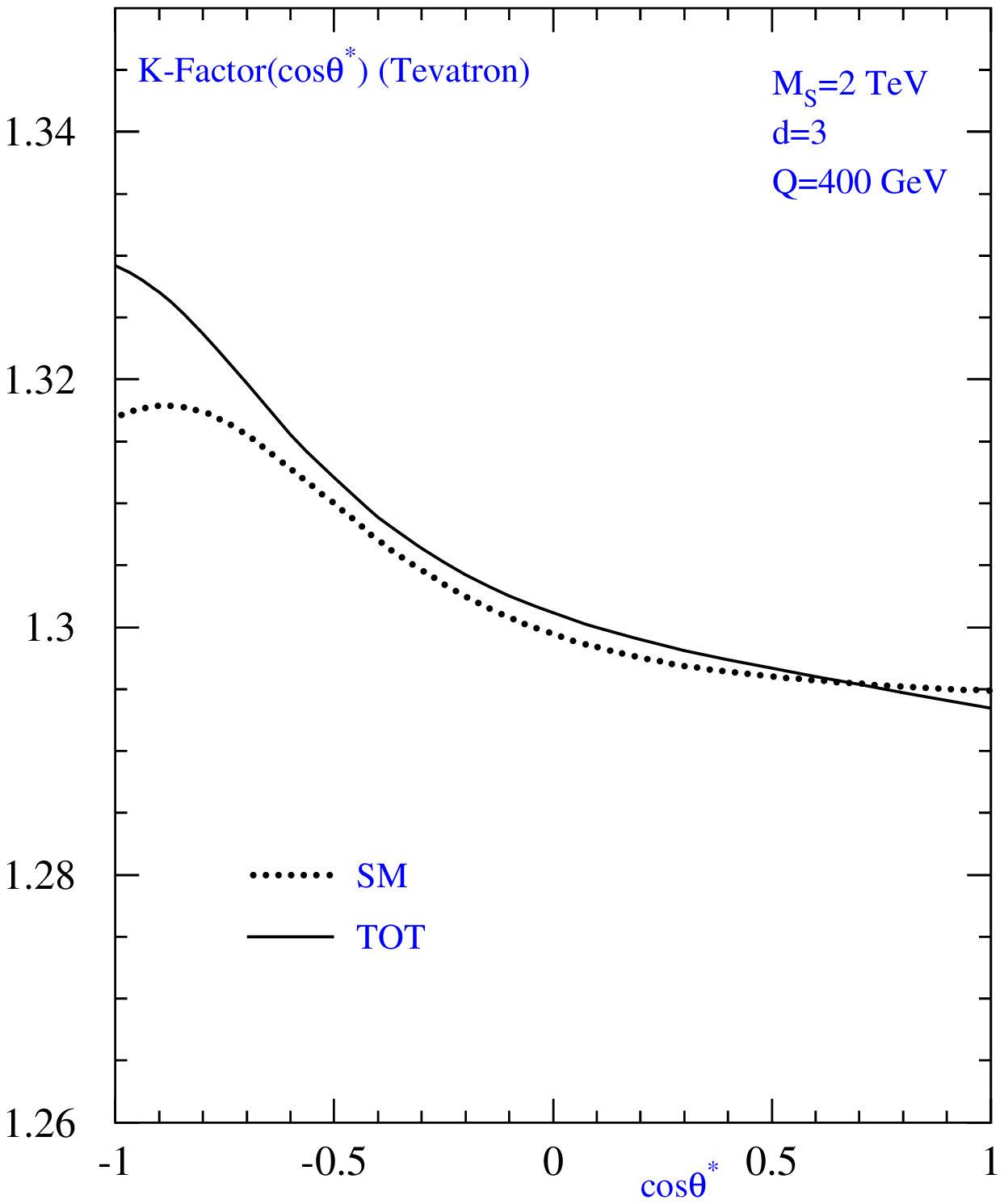,width=15cm,height=16cm,angle=0}}
\vspace{5mm}
\centerline{\bf Fig.~3b}
\end{figure}
                                                                                
\eject
\begin{figure}[htb]
\vspace{1mm}
\centerline{\epsfig{file=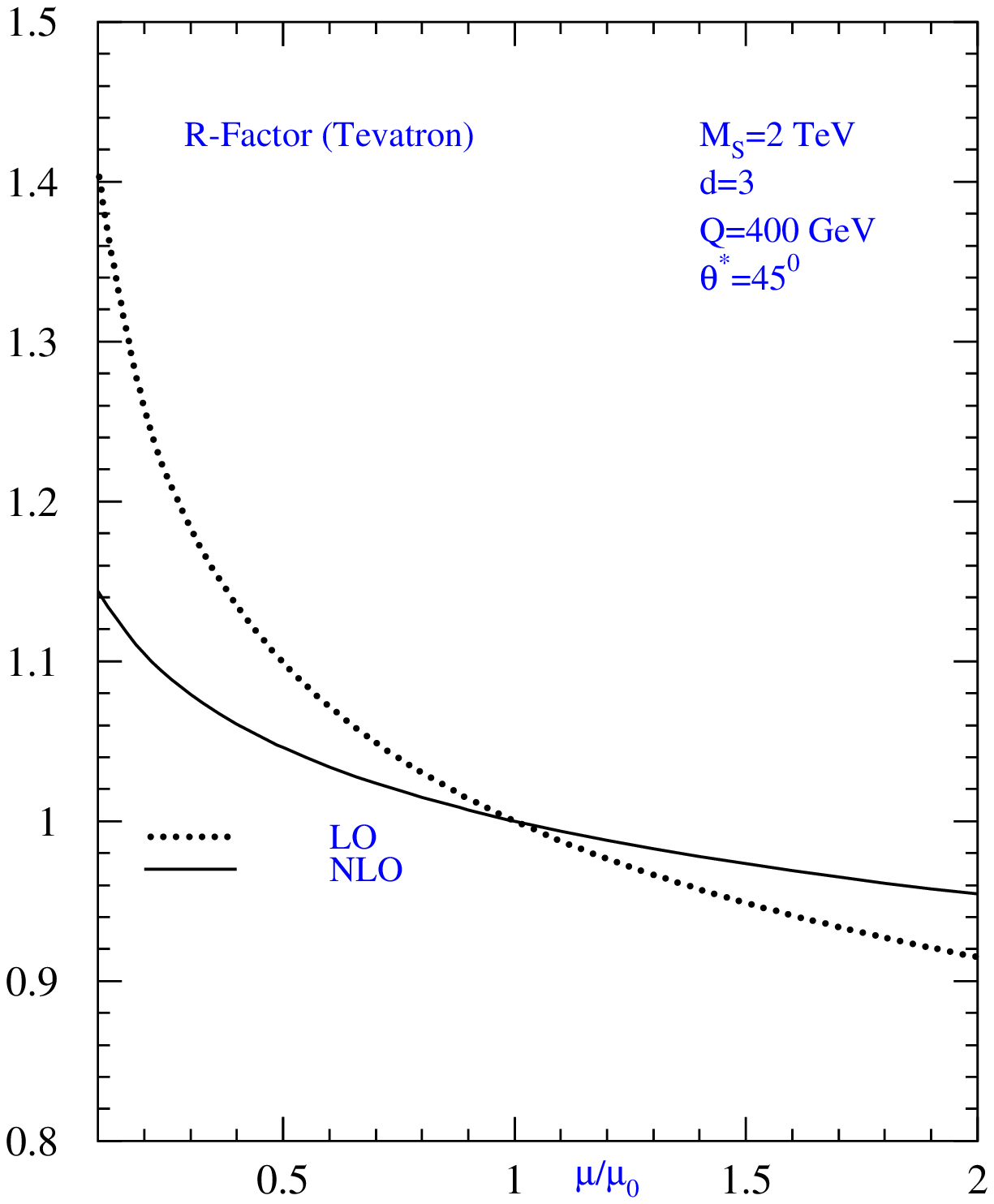,width=15cm,height=16cm,angle=0}}
\vspace{5mm}
\centerline{\bf Fig.~3c}
\end{figure}
                                                                                
\eject
\begin{figure}[htb]
\vspace{1mm}
\centerline{\epsfig{file=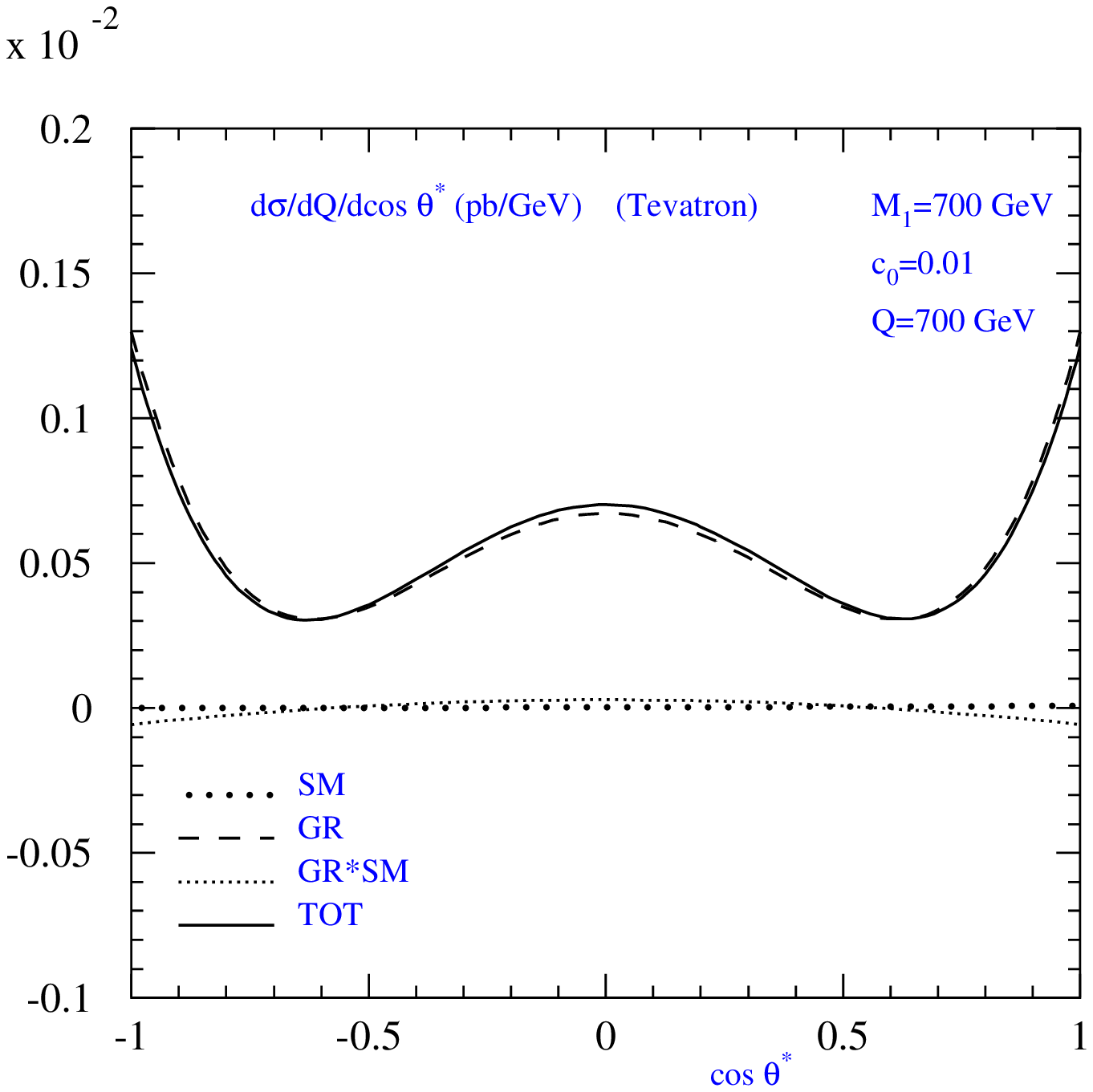,width=15cm,height=16cm,angle=0}}
\vspace{5mm}
\centerline{\bf Fig.~4a}
\end{figure}
                                                                                
\eject
\begin{figure}[htb]
\vspace{1mm}
\centerline{\epsfig{file=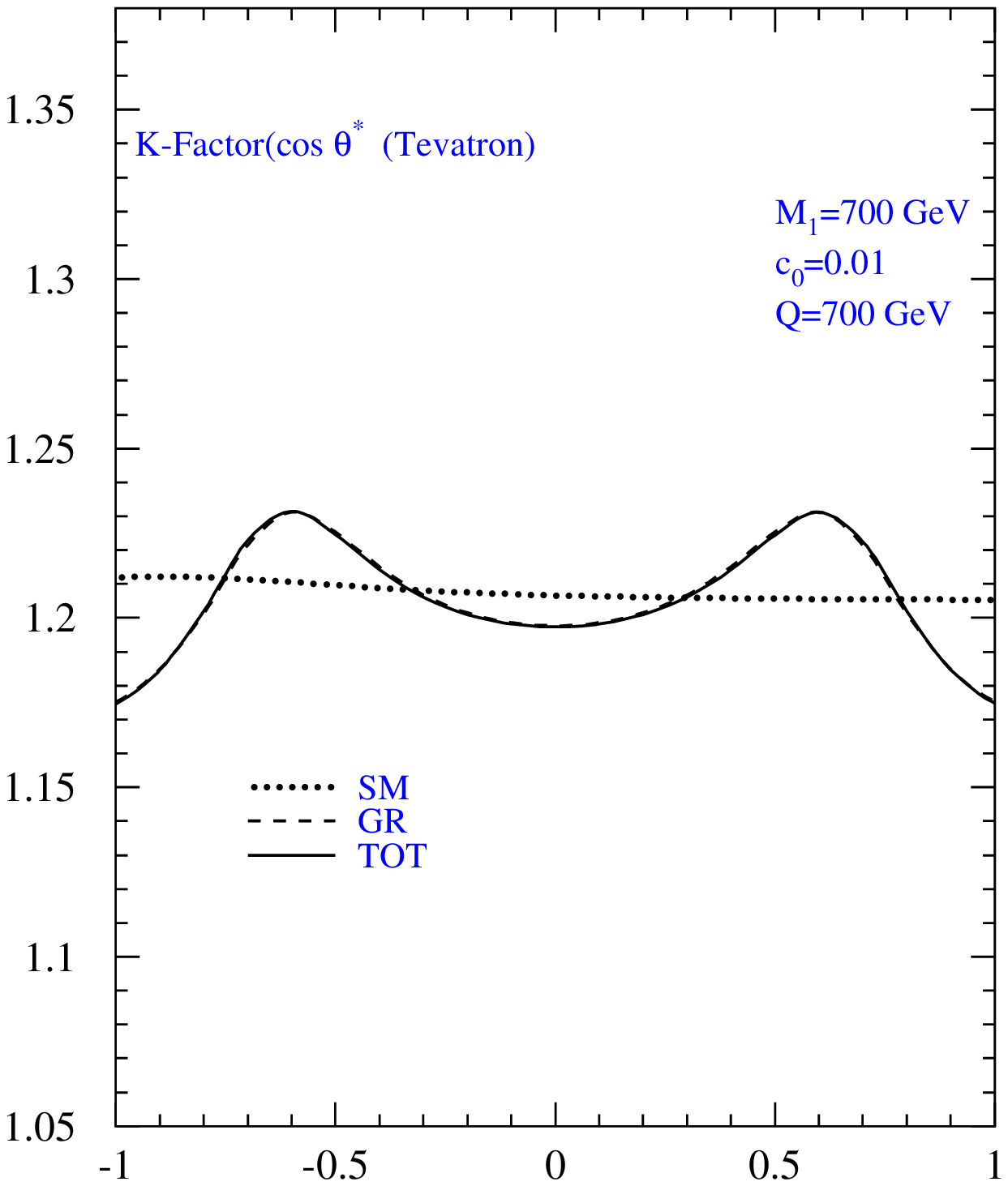,width=15cm,height=16cm,angle=0}}
\vspace{5mm}
\centerline{\bf Fig.~4b}
\end{figure}
                                                                                
\eject
\begin{figure}[htb]
\vspace{1mm}
\centerline{\epsfig{file=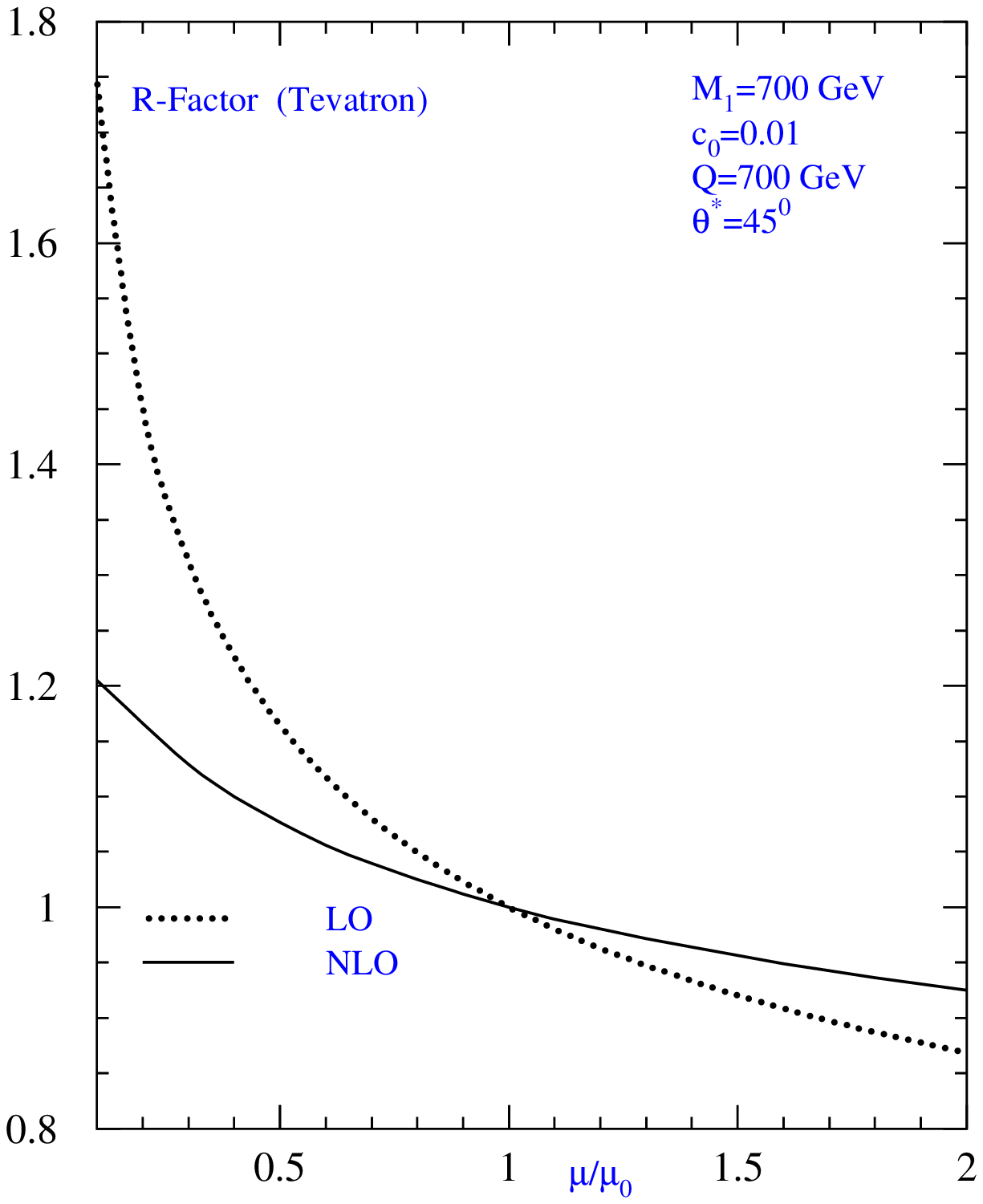,width=15cm,height=16cm,angle=0}}
\vspace{5mm}
\centerline{\bf Fig.~4c}
\end{figure}


\end{document}